International Conference
on Industrial Engineering and Systems Management

**IESM' 2009**

May 13 - 15, 2009

MONTREAL - CANADA

# A decision support system for manufacturing improvement and relocation prevention in Thailand: Supply Chain perspective[*]

Napaporn REEVEERAKUL[a], Ridha DERROUICHE[a],
Nopasit CHAKPITAK[a], Yacine OUZROUT[b],
Napat HARNPORNCHAI[a], Abdelaziz BOURAS[b]

[a]*CAMT, Chiang Mai University, Thailand*

[b]*LIESP, Université de Lyon (Lumière Lyon 2), France*

*Abstract*

The low economic growth and competition among neighbouring countries has caused Foreign Direct Investments (FDIs) to relocate their businesses. In order to prevent further business relocation, this paper proposes an integrated framework based on the supply chain to help analyse decision making for plant situations and enhance manufacturing performance. The context of this perspective is applied to manufacturers located in the industrial estate region of Lumphun province, Thailand. Data collection and review of literature was used to identify the factors that influence industrial investment. The SCOR model was used to define the parameters, which were then used in Arena simulation. The simulation needs to describe factors affected on industrial performance. From this result, an integrated analysis model was built and the importance of supply chain collaboration was identified. A Multi Agent System (MAS) was suggested to enhance of the effective interaction between supply chain' agents. It is in order to mitigate risks among them.

## 1 Introduction

Thailand is classified as a developing country. The majority of Thai people work in the agriculture sector but their sectors have accrued faster to more closely link to the international economy, for example; jobs in manufacturing for export or labouring jobs in companies. Those who remained in the domestic economy, such as small scale farmers, generally received fewer benefits, proportionately [1]. In electronics manufacturing, high technology and cost investment, skill requirement, continuous research and development, and intensive labour

---

are needed. Most electronics manufacturers depends on Foreign Direct Invesments (FDIs) with both funding and high technology. At the same time, FDIs are interested in investment in developing countries in order to get the benefits of low labour costs.

Particularly, Lumphun province is the location of a large industrial estate in the northern part of Thailand. Earnings from industrial labor is a key factor driving the province's growth. Sixty-three factories are located in this area, most manufacturing products of which 31.16% are manufacturing electronics. 43,456 persons work in Lumphun province [2], while the workforce in the electronics factories in the Lumphun estate represents 83.33% of the manufacturing workforce in this province [3]. However, it has been noted that plant closures have increased to 4.76% during the period 2005 to 2007 [2]. Low local economic growth and competition with neighbouring countries are the main reasons, leading to business relocation to countries with cheaper labour cost. These unwelcome circumstances lead to economic problems in the province, affecting income and finally leading to social problems.

In this context, we are proposing here an analysis of manufacturers' decisions on plant situations. In order to prevent relocation of plants, the model studies three perspectives: finance, supply chain and knowledge skill within companies. However, the main perspective discussed here is the supply chain analysis.

In the next section we summarize the review of literature about the factors influenced decision on plant situation. In section 3, we introduce a new framework on making a decision on plant situation. The framework helps manufacturer to analyse and make a decision on plant's situation, based on a three perspectives with Finance, Supply chain and Human skill. As an illustration, we apply this framework to case study with supply chain simulation technique in section 4, and we show how this supply chain collaboration important for manufacturers. Finally, we introduce in the section 5, a Multi Agent System (MAS) needs to enhance of the effective interaction between supply chain' agents through an agent communication language.

## 2 Context and previous works

Recent studies have proposed some directions to improve the decision making in Supply Chain.

### 2.1 Context of this work: Supply Chain Collaboration

Supply chain collaboration is often defined as two or more enterprises working together to create a competitive advantage and higher profits that cannot be achieved by acting alone [4]. Supply Chain Council, [5] proposes a structure of a supply chain around five distinct management processes as shown in Figure 1.

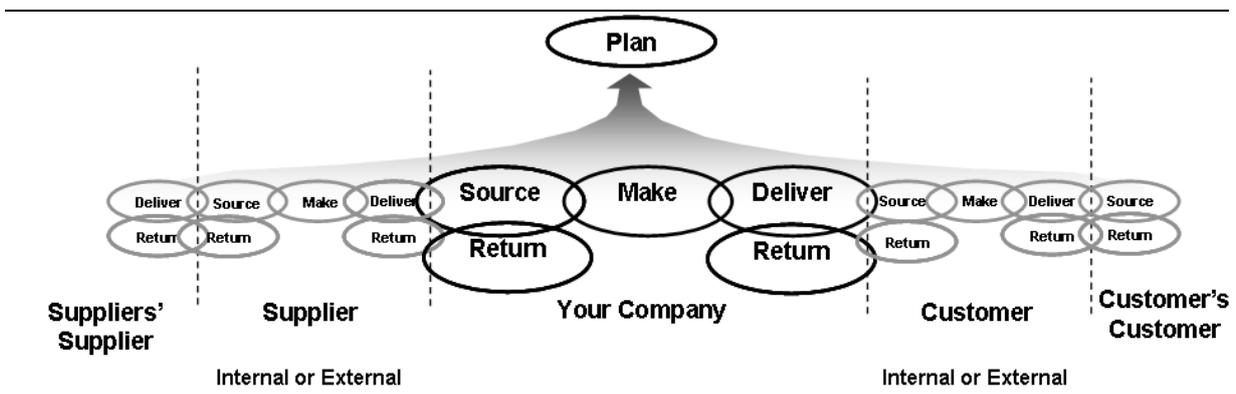

Fig.1. Supply chain structure according to SCOR-M [5]

The advent of supply chain collaboration creates the need, at intra-enterprise level (between the different plants and the different processes) and at inter-enterprises level (between the different partners) to pay special attention to the understanding of collaboration in order to prepare the partners to create collaborative efforts successfully [6]. Supply chain collaboration implies that all factors relevant need to be taken into account to produce a good decision.

### 2.2 Review of literature: Factors influencing decisions on plant situation

Chan et al., [7] found that the main reasons for business relocations were cost savings and business expansion, regardless of whether the firm was a plant or a headquarters. Other literature from Qi Chun, Jhon C.S. Tang,[8] proposed a genetic algorithm (GA) as an analytical tool to define fitness function as a variable selection algorithm and analysis of inward Foreign Direct Investment (FDI). They found 24 independent variables that affect FDI in China which the most often selected determinants of FDI flows into industry are productivity, research and development investment, labour investment and educational level. So far as investment in manufacturing is concerned profit which is a key factor for the investor. Minimizing costs while maximizing revenue is undertaken to maximize profit. For financial measurement, net present value (NPV) is one well-known strategy that ascertains the time value of money invested in a business. Generally, the approach used for optimize decision making. A number of production planning studies have subsequently been evolved with the aim of maximizing the NPV [9].

In recent years, many companies have observed that beside their "traditional" risks arising from their business activities, new risks emerge from sources that are often related to the close collaboration within their supply chain networks [9]. Supply-Chain Council (SCC) Risk Management Team & Douglas Kent (December,2007),[10], argued that today's organizations recognize that "preservation of shareholder value" is highly important in supply chain management. So disruption in the supply chain will impact on the company's overall performance.

Indeed, several authors developed this question under various points of view such as: to make a decision on plant relocation from Europe to China, incorporate management participation method was used by considering return on profitability and soft variables, example of: loyalty skill and local performance [11]. Besides, N.Viswanadham and Kannan Balaji [12] proposed quantitative model for optimal decision between FDI and outsourcing with multi-stages. At various stages of supply chain risk due to supply chain, inventory and transportation cost. To investigate the fact of technology transfer by FDI in electronics sector in South Africa, the researchers criticized the need to be integrated from start to acquire capabilities focused on developing knowledge based, security and subcontract existence [13]. The table 1 synthesizes and classifies the various studies according to the attributes used.

Table 1 Factor analysis on plant allocation and investment decision from review of literature

| *Factor analysis on plant allocation/investment decision* | [3] | [8] | [11] | [12] | [13] | [14] | [17] | Total |
|---|---|---|---|---|---|---|---|---|
| 1. Supply chain cost | x | | x | x | | | | **3** |
| 2. Production quality | | x | | | | | | 1 |
| 3. Market demand | | | | x | x | x | | **3** |
| 4. Logistics cost | | | | x | | | | 1 |
| 5. Employee skills | | | x | | x | | | 2 |
| 6. Expatriate management | | | x | | | | | 1 |
| 7. Technological support | x | | | | | | | 1 |
| 8. Loyalty to the employer | | | x | | | | | 1 |
| 9. Labour productivity | | | x | | | | | 1 |
| 10. Close to supplier (Lead time) | x | | | x | | x | | **3** |
| 11. Economic development policy | x | | | x | | x | | **3** |
| 12. In-bounded lead time | | | | x | | | | 1 |
| 13. Supply chain risk | | | | x | | x | x | **3** |
| 14. No. of project in R&D | x | | | | | | x | 2 |
| 15. Total investment in R&D | | x | | | | | | 1 |
| 16. Return on asset | | x | | | | | | 1 |
| 17. Internal culture | | | | | | x | | 1 |
| 18. Competition | x | | | | | x | | 2 |
| 19. Firm size | | | | | | | x | 1 |
| 20. Capital to Labour | | x | | | | | | 1 |
| 21. Total sales values | | x | | | | | | 1 |
| 22. Export sales values | | x | | | | | | 1 |
| 23. No. of high educated employee | | x | | | | | | 1 |
| 24. Wages of domestic employees | | x | | | | | | 1 |
| 25. Security | | | | | x | | | 1 |
| 26. Subcontractors existence | | | | | x | | | 1 |

From a review of literature, it is clear that there are many factors affecting plant relocation. Most researchers are more focused on cost of production, supply chain between supplier and customer, for instance, market demand and supplier lead time or even supply chain risk. However, the consideration on external factors as economic development policy cannot be neglected.

*2.3 Data collection by questionnaire*

To confirm and assure factors that influenced plant relocation and investment decisions in Lumphun industrial estate area, research was done by use of questionnaire. Manufacturing businesses which received questionnaires in the Lumphun industrial estate, northern Thailand, included electronics, jewellery, mechanical parts and components, and garment industries. There were two sorts of questionnaires the first, to examine factors influencing investment decisions. The second, searched for the causes leading to plants closing, relocating or expanding business. The results based on factors influencing investment decisions showed 3 areas:

1) The need for proximity of a calibration centre: this is such an important factor, even though most manufacturers had their own research and development centre located on their headquarter site. However a research and development centre is moderately required for manufacturers.

2) Transportation: according to outbound logistics, land and air freight are the most commonly used transportation methods.

3) Competency and skill based requirement: technical supporting into the real workplace is strongly needed factors to improve knowledge worker's performance by manufacturer.

In addition, the need for telecommunication such as leased line, Asymmetric Digital Subscriber Line (ADSL) or Wireless Local Area Network (WLAN) and supporting infrastructure cannot disregarded.

Besides the second questionnaire showed that, different factors impacted on relocations, expanding and closing of plants. Lack of working skills and abilities are a strong factor causing plant closure, while low labour costs and under-developed infrastructure are shown as the dominant criteria impacting on decisions to relocate and expand businesses. Nevertheless, external factors such as the strong exchange rate of the Thai baht, political uncertainty and disadvantages of regulations and policy are all relatively key factors as shown in Figure 2.

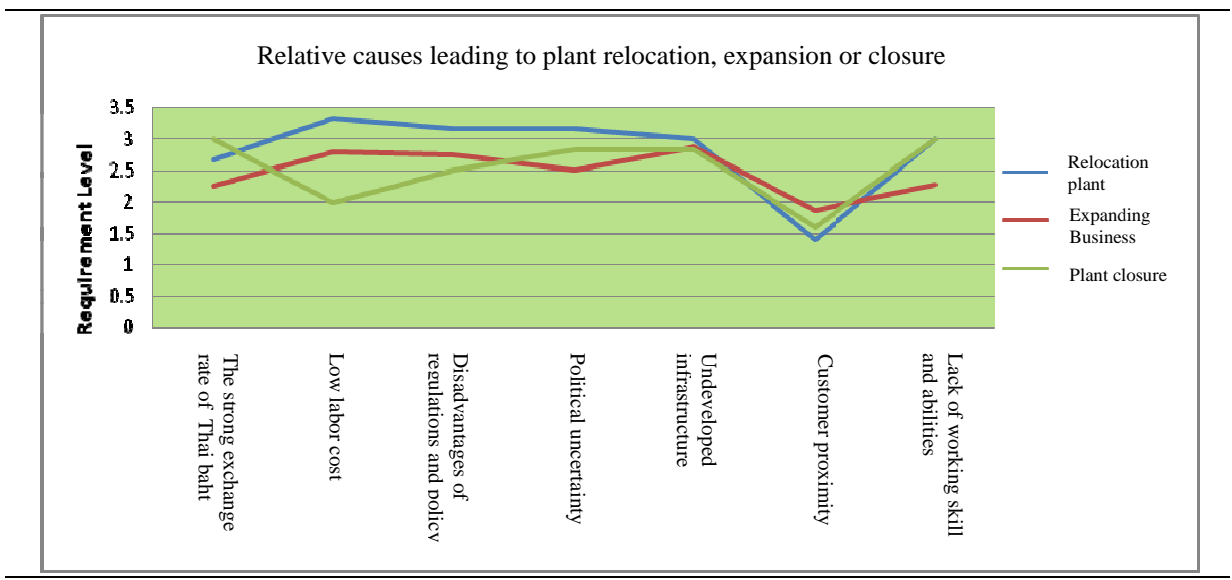

Fig. 2. Comparisons of on relative causes with 3 scenarios

According to the analysis of literature and the survey, it is obvious that the majority of research on plant investment decisions do not agree on a unique set of criteria (attributes) to typify relations between relocation decisions and the influencing factors. The attributes used in each research are different and the contents of each type of decision remain sometimes very general, such as the variables affecting FDI based on specific production or transportation costs in the supply chain. Some other authors detailed the attributes which influence a decision in each type of situation but their studies were focused on specific orientations, for instance, supplies existence [3],[12],[14], Supply chain risk [12],[14],[17], even the criteria on economic development policy [3],[12],[14], on supply chain cost [3],[11],[12], and research and development support [3],[17] which cannot be generalized. We can identify influencing factors on plant decisions, based on three main necessities. These are supply chain, human skill and financial consideration.

On the basis of these conclusions, we propose integrated framework to help decisions on plant situations and prevent relocation by the improvement of manufacturers' performance.

## 3   A framework of decision making and prevention of plant relocation: Supply Chain context

According to the previous analysis, a framework for making decisions on plant situations was introduced. The framework helps the manufacturer to analyse and make a decision on a plant's situation. Figure.3 shows the framework explanation based on three perspectives.

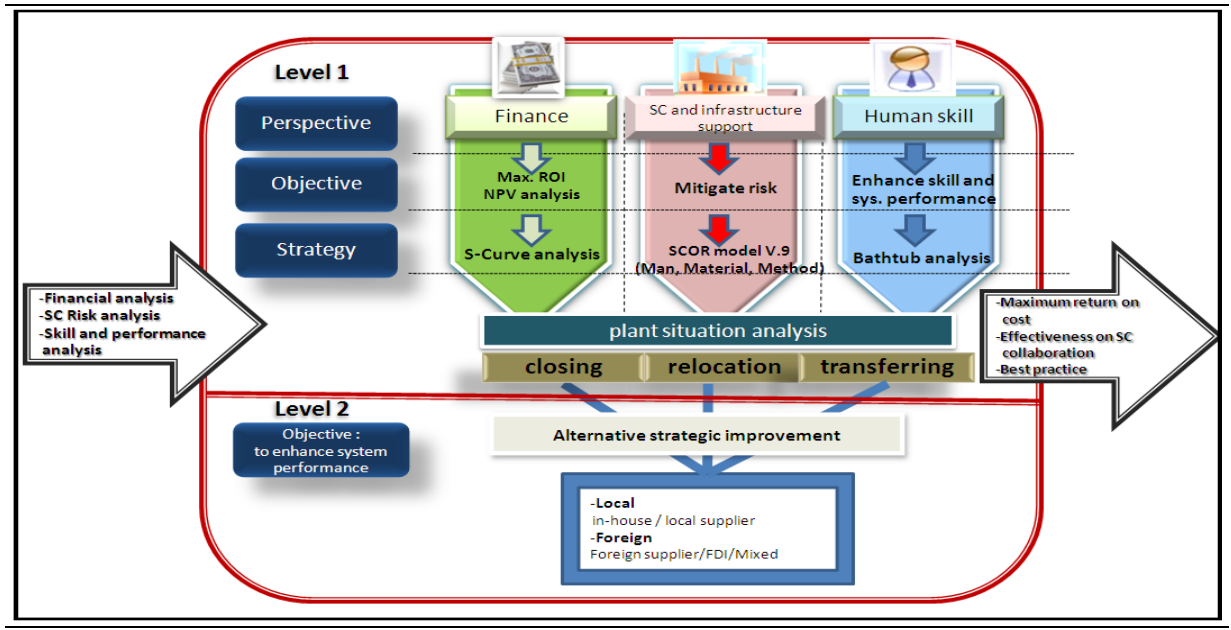

Fig. 3. The Framework decision on plant situation and system performance improvement

On supply chain management, in order to improve manufacturing performance, it is very necessary for firms to enhance effectiveness on collaboration, the integration of supply chain, mitigate risk and, provide more benefits to achieve customer satisfaction. Moreover, manufactures can keep and extend their operational life. To avoid human failure, skill and performance improvements are needed to develop manufacturing systems. Knowledge sharing among knowledge workers and asset transferring are useful approaches to eventually extend a factory's operation. Besides, according to investment in manufacturing, the consideration on financial aspect is implemented by s-curve analysis. The maximum expectation on return on investment is an objective function for investors.

Focussing on the three perspectives, the investor is interested in maximum return on investment (ROI). Factory objective is to mitigate supply chain risks. Meanwhile, enhanced skill and worker performance helps to reduce human error in the working area. Figure.3 describes the explanation of the objective to be successful and the strategic decision on plant situation and system performance improvement. As shown in the framework, to prevent unwelcome situations, alternative strategies on supply chain management by enhancing system performance and their collaboration are adopted. So this paper aims at supply chain perspective.

Nowadays, companies face many critical challenges with regards to supply chain management. One of the main challenges arises out of trends associated with lean management. Strongly synchronized interfaces and the reduction of inventory cause a greater dependence on the supply chain partners. Close co-operation between supply chain partners exacerbates the risk exposure of companies in a supply chain [9]. To improve manufacturing industry, the integration of supply chain and risk analysis is necessary. Besides, supply chain analysis is used to identify the impact of changes in the location and/or capacity of facilities on supply, manufacturing and distribution activities. This impact can be measured by determining the required input quantities (materials, labour, capital and energy) and the respective unit prices for alternative locations and capacity level. In addition, factors in the international environment like taxes, barriers to trade and exchange rate, will affect allocation strategies [11].

As mentioned above, most arguments are explained in qualitative aspects. Then the practical data is used for an implementation in quantitative aspect. The application on simulation model based on SCOR (Supply Chain Operations Reference) model analysis was used.

Therefore model with SCOR on level 1 related to scope of processes (Plan, Source, Make, Delivery and Return) and processes categories, process element information inputs, outputs and activities which defines the ability to compete operations strategy. To describe unambiguously and communicate the process, a framework of relationships among the standard processes in supply chain was explained with the SCOR-model. The SCOR-model captures the Supply Chain Council's consensus view of supply chain management. While much of the underlying content of the model has been used by the practitioner for many years, the SCOR-model provides a unique framework that links business process, metrics, best practices and technology features into an effective supply chain management and related supply chain improvement activities [5]. SCOR-Model used for performance attributes on supply chain simulation as shown in Table 2.

Table 2 Performance attributes and metrics used for the simulation model

| *Performance attribute* | *Metric* |
| --- | --- |
| Reliability | Delivery performance |
| | Fill rate |
| | Perfect order fulfilment |
| Responsiveness | Order fulfilment lead time |
| Flexible | Demand chain response time |
| | Delivery flexibility |
| Cost | SC management cost |
| | Cost of goods sold |
| | Value added productivity |
| Asset | Cash to Cash cycle time |
| | Inventory days of supply |
| | Asset Turns |

## 4   Application of the framework: Case of Lumphun industrial estate, Thailand

To specify the supply chain system in a manufacturing company, a flow diagram helps to illustrate the system [Figure.4]. The flows are used to model the relation between variables in a continuous system.

A commonplace situation in manufacturing is one supply source providing material to manufacturer, one distributor receiving finished goods and requesting order. Goods are made and products are made at customer request. Turnover rate on worker recruitment and inventory level was also considered before releasing work order to production line, in which 10 workers used for production.

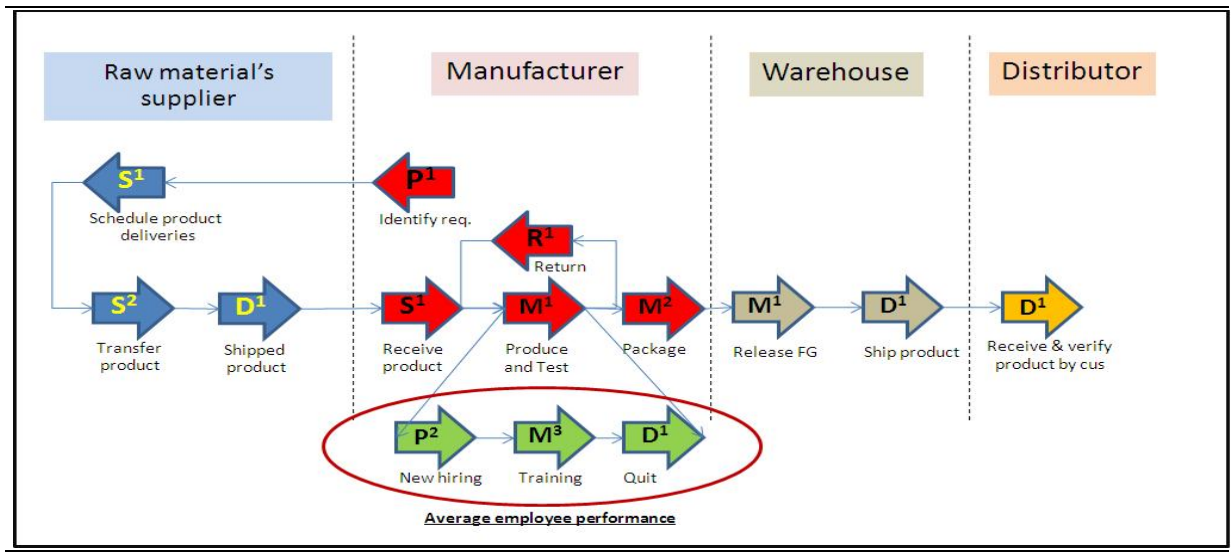

Fig. 4. Flow diagram on manufacturing simulation model

In order to describe the interaction between relevant parameters and define critical factors affected to system performance, simulation technique using Arena software is implemented. As mentioned above, according to simple flow, in which complies with SCOR-model as shown leads to supply chain simulation. The illustration on Arena model is presented. The simulation was run by using assigned variables as the system's performance. The simulation required comparison on system performance such as output productivity, cycle time, while variables were adjusted. The adjustment is applied as an approach, in order to describe the most effect on supply chain and system performance. The results represent sensitivity analysis. It has been noted that increase in turnover was distinctly affected by cycle time and waiting time in process. Meanwhile, time between process and amount of work in process are directly alternated according to demand fluctuation rate. Moreover, comparing with other criteria, rejected parts or loses in production directly result from human error and skill deficit. For example, increasing turnover rate by 32 % impacts the cycle time with a 3.57% increase. Moreover, demand interruption in every UNIF (10, 15) days, caused a 31.31% increase in worker utilization. While work in process (WIP) quantity and waiting time in process are also obviously increased. More results were explained as presented in table 3. As shown in table 3, on an inbound supply chain, the change in high percentage of turnover, lack of workers' skill and effectiveness impact distinctly on cycle time expansion and losing time in production. On the other hand, an outbound supply chain with variation on demand and supplier lead time affected in processing time and quantities in process. This can be assured that the important of management and collaboration in supply chain helps to enhance manufacturing performance and need for business expansion.

Table 3 Analysis result by factors affected on system performance

| SCOR Card | | Issue | Measurement | | | | |
|---|---|---|---|---|---|---|---|
| Performance attribute | SCOR Level 1 Metrics | | Cycle time | Worker utilization | Waiting time in process | No. WIP | Rejected part in prod. |
| Asset | Resource turns | Turnover (32% increased) | *+3.57%* | *Not changed* | *+0.0766 %* | *Not changed* | *N/A* |
| Supply chain reliability | Fill rate | Demand fluctuation rate (Interruption 20 units every UNIF ( 10 , 15) days ) | *N/A* | *+31.31%* | *+30.96%* | *+22.76%* | *N/A* |
| Cost: | Value added employee productivity | Human error (Increased 5% | *+4.82%* | *+4.3%* | *+0.256%* | *Not changed* | *+102.94 %* |
| Responsive-ness | Order fulfillment lead time | Supplier lead time (1day to 7 days) | *+1.16%* | *-0.1243%* | *Not changed* | *+2.414%* | *N/A* |
| Flexible | Supply chain response time | | | | | | |

## 5   Toward a Multi-Agent System Model

A Multi Agents System (MAS) is the system in which a number of intelligent agents interact with each other. Each agent has their own parameters, indicators and decisions to express their status [15]. As shown in Figure.5, a simple supply chain agent comprises supplier, manufacturer, distributor and customer.

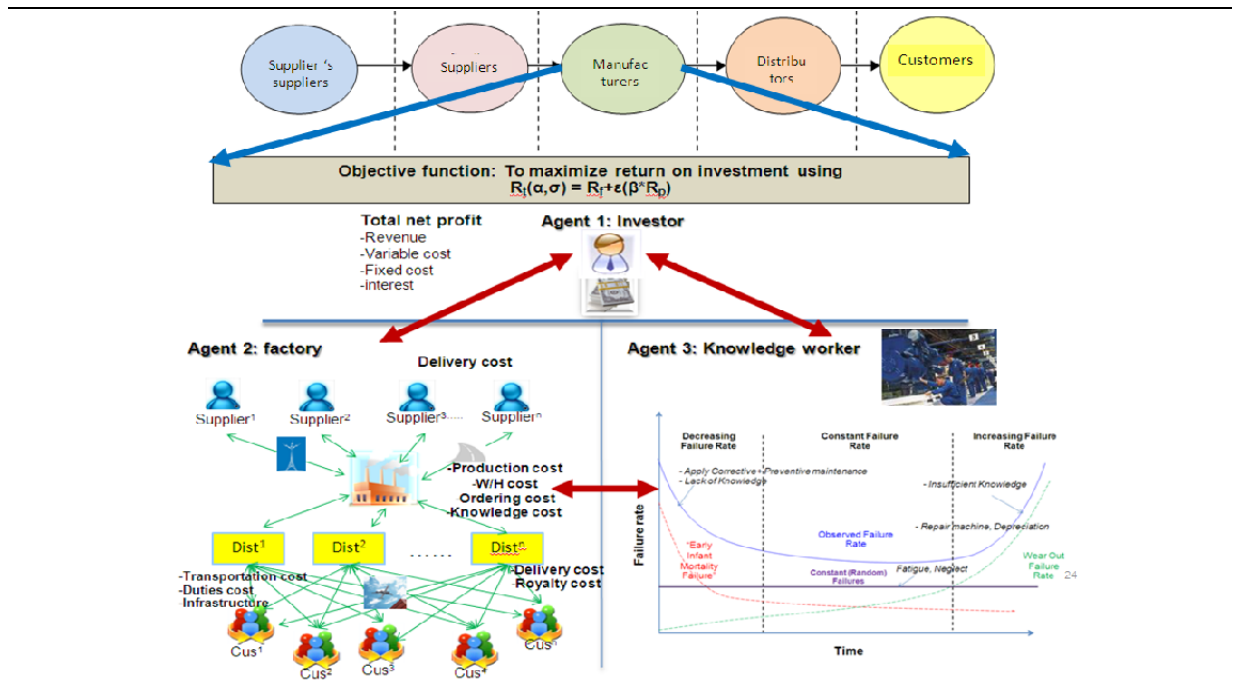

Fig. 5. Supply chain decision on plant situation

Consequently, supply chain analysis will be applied to decision's model framework. The analysis on collaboration among each partner in supply chain network is mainly focused. For further research, in order to enhance of the effective interaction between supply chain' agents through an agent communication language, a Multi Agents System approach will be used. Certainly, it is beneficial for a firm to develop risk mitigation strategies proactively to handle potential disruptions before risk occur [16].

## 6  Conclusion

To support a decision to prevent plant relocation, a proposed framework on manufacturers' decisions on plant situations was introduced, focusing on three perspectives on the situation: Finance, Supply chain and Human factors. The recent study focused on Suppy Chain. In this context, qualitative and quantitative analysis by survey using questionnaires and a model on Arena simulation were implemented. The survey showed factors that influenced investment in three areas: the need for a calibration centre, transportation and worker skill. The strongly impacting factor leading to relocation related to infrastructure and cost, while lack of working skill and abilities caused plant closure situation. Besides, simulation based on SCOR-model supported results on inbound and outbound supply chain, are important on management and collaborate in chain networking. This helps to enhance performance and expand their businesses.

The introduction of Multi Agent System (MAS) is used to simulate the collaboration in supply chain network. In future work an integrated analysis on finance with supply chain risk will lead to making a decision on plant's situation. Nevertheless, enhancing skill and system performance, and knowledge management cannot be abandoned.